\documentstyle[12pt,aasms4]{article}
\slugcomment{To be published in The 
Astrophysical Journal (Letters)}
\lefthead{Nemiroff et al.}
\righthead{GRB Spikes}

\begin{document}

\title{GRB Spikes Could Resolve Stars}

\author{Robert J. Nemiroff}
\affil{Department of Physics, Michigan 
Technological University,
Houghton, MI  49931}
\author{Jay P. Norris, Jerry T. 
Bonnell\altaffilmark{1} and}
\affil{NASA Goddard Space Flight Center, 
Code 660.1, Greenbelt, MD, 20771}
\author{Gabriela F. Marani}
\affil{Center for Earth and Space Research, 
George Mason University, Fairfax, 
VA  22030}
\altaffiltext{1}{Universities Space Research 
Association}

\begin{abstract}
GRBs vary more rapidly than any other known
cosmological phenomena.  The lower limits
of this variability have not yet been explored.
Improvements in detectors would reveal or limit the 
actual rate of short GRBs.
Were microsecond ``spike" GRBs to exist and be detectable,
they would time-resolve stellar mass objects throughout the universe
by their gravitational microlensing effect.  Analyzing the 
time structure of sufficient numbers of GRB spikes 
would reveal or limit $\Omega_{star}$, $\Omega_{MACHO}$, 
and/or $\Omega_{baryon}$.
\end{abstract}


\section{Introduction}

The duration and peak flux distributions of gamma 
ray bursts (GRBs) have been measured and 
discussed continually since discovery (Klebesadel, 
Strong and Olson 1973).  Since 1991 these 
distributions have been best sampled by the 
relatively sensitive Burst and Transient Source 
Detector (BATSE) onboard the Compton Gamma 
Ray Observatory (CGRO).  Intrinsic detection 
thresholds, however, place fundamental 
limits on the duration and peak fluxes of GRBs BATSE 
can detect.  BATSE's onboard trigger criteria make 
it most sensitive to GRBs with durations of about 1 
second and peak fluxes above 1 photon cm$^{-2}$ 
sec$^{-1}$ (Fishman et al. 1994a; Meegan et al. 
1997).  Dimmer and/or shorter bursts, however, 
may carry significant cosmological utility.

The GRB duration distribution has been 
studied previously mostly with regard to its 
unusual bi-modal appearance (Cline \& Desai 1974; 
Norris et al. 1984; Hurley et al. 1991; Kouveliotou 
et al. 1993).  In this paper we will focus more 
closely on extrapolating the BATSE 3B 
duration distribution to durations well below the 
minimum time scale of BATSE's trigger 
accumulation time: 64 ms.  We will implicitly 
assume that the majority of 
GRBs occur at cosmological distances for reasons
outlined by Paczynski (1995).  

The possibility of gravitational lensing of GRBs was
first discussed by Paczynski (1987), where the 
detection of clusters of microlensing images 
was considered.  The possibility of microlensing differentially
affecting macro-images created by the lensing of intermediate 
galaxies was discussed by Nemiroff et al. (1994).  
Williams \& Wijers (1997) showed that in the presence of 
significant lensing optical depth and shear, the time delay
between GRB images could be as long as several 
milliseconds.  Searches for longer duration
gravitational lens effects in GRB data have been carried 
out by Nemiroff et al. (1993, 1994) and Marani et al. (1998).

During BATSE's first three years (comprising the time
of the BATSE 3B catalog: Meegan et 
al. 1997), BATSE's onboard trigger criteria were 
relatively constant.  On average, BATSE was 
trigger sensitive to any point on the sky about 48 \% 
of the time, between 50 and 300 keV on 64 ms, 256 
ms, and 1024 ms time scales (Meegan et al. 1997).  
Peaks in the time series greater than 5.5 $\sigma$ 
over a previous 17 second background triggered the 
instrument into burst mode, where more detailed 
time information about the burst was recorded.

These very detection thresholds, however, limit 
which GRBs BATSE can detect.  In particular, 
GRBs with durations below the BATSE 64 ms 
trigger threshold are only detected by BATSE when 
accumulated flux over 64 ms is sufficiently large.  
Theoretically, a GRB with arbitrarily high peak flux 
but correspondingly short duration would go 
undetected by BATSE.  Practically, BATSE is 
increasingly {\it in}sensitive to shorter GRBs, 
demanding they have increasingly high peak flux 
for detection (Norris et al. 1984; Fishman et al. 1994a; 
Lee \& Petrosian 1996).  In particular, 
Lee \& Petrosian (1996) model these effects 
as a bias against the detection of bursts with durations 
below the minimum 64 ms trigger time scale, 
extrapolating a best fit on the rate down to a duration of 
10 ms.

In \S 2 the extent to which BATSE
3B trigger scheme limits the rate of extremely short 
duration GRBs is explored.
In contrast to Lee \& Petrosian (1996), we are more interested
in what rates are reasonably allowed for short GRBs rather 
than in finding the best fit rate.  In \S 3 the potential 
utility and detectability of GRB spikes is discussed. 
In \S 4 some discussion and conclusions 
are given.

\section{Extrapolations to Short Duration}

To explore possibilities on how many short 
duration GRBs could exist and go undetected by
BATSE, we first focus on how BATSE's internal
trigger creates a practical relation between 
peak flux and duration.  
Figure 1 shows a plot of duration versus peak flux 
for BATSE 3B GRBs.  Only those bursts with listed 
$T_{50}$ and $T_{90}$ durations, and peak fluxes 
on all three trigger time scales (64, 256, and 1024 
ms) are plotted: 807 in all.  This sample was chosen
because it is relatively well understood and 
could be tested for a variety of potentially confounding 
internal correlations.  BATSE altered its 
triggering criteria immediately following the
conclusion of 3B catalog data collection, so even
though more BATSE GRBs exist at the time of 
this writing, to include them would introduce an
inhomogeneity.  

On the x-axis an estimate of duration was plotted: 
$T_{50}$ from the BATSE 3B catalog, the time 
between when 25\% and 75\% of the total GRB 
counts were accumulated.  This duration measure is 
known to be biased in the sense that low peak flux 
GRBs have at least slightly shorter BATSE 
$T_{50}$s, on the average, than if they had higher 
peak flux (Bonnell et al. 1996; Koshut et al. 1996), 
even though $T_{50}$ is designed to be 
independent of peak flux (Kouveliotou et al. 1993; 
Lee \& Petrosian 1996).

On the y-axis an estimate of true peak flux was 
plotted.  Specifically, the published BATSE peak flux 
estimate on the 64 ms time scale was used.
A dotted line at 64 ms is drawn in vertically on 
Figure 1 to indicate BATSE's minimum trigger time 
scale.  

The solid line drawn on Figure 1 is a rough 
indicator of detection limits, with the region below 
delineated by hashed lines running from the lower 
left to the upper right.  This region, labeled 
``BATSE Untriggered", is generally invisible to 
BATSE's onboard trigger criteria.  There is at least 
a 70 \% chance that a GRB in this region would go 
untriggered by BATSE, according to the exposure versus
peak flux table published with the BATSE 1B catalog  
(Fishman et al. 1994a).  At 64 ms, this solid 
line lies at a peak flux level of about 
1 photon cm$^{-2}$ sec$^{-1}$.

At durations shorter than 64 ms, the solid line rises 
toward higher peak fluxes, indicating BATSE's 
insensitivity to shorter bursts.  The approximate 
detection threshold is estimated from trigger criteria 
to be 
 \begin{equation}
        P(t < 64 \ {\rm ms}) \sim P(64 \ {\rm ms}) 
                       {{64 \ {\rm ms}}\over
                         {t_{dur}{\rm (ms)}}},
 \end{equation}
where $t_{dur}$ is a duration estimate of the GRB.
This insensitivity is caused solely by BATSE's 64 ms
trigger criteria.  

Even if a trigger bin size below 64 ms could somehow 
be enabled, BATSE's background would then limit the 
minimum measurable peak
flux at a given $t_{dur}$. 
The decreased sensitivity is a direct result of the 
decreased number of counts available from these 
GRBs, given that the trigger depends on some
measured $\sigma$ over background, and that
$\sigma$ scales as the square root of the background level.  
For a constant $\sigma$ trigger, 
 \begin{equation}
 P(t < 64 \ {\rm ms}) \sim P(64 \ {\rm ms}) 
                \left(      {{64 \ {\rm ms}}\over
                         {t_{dur}{\rm (ms)}}} \right)^{1/2} .
\end{equation}
This background limit is shown by the dashed line in Figure 1.

Although equations 1 and 2 have similar terms, they 
demonstrate different things.  Equation 1 
describes how intense the peak flux of a spike GRB
must be to trigger given a constant bin size.  
Equation 2, on the other hand, shows how high the peak
flux of a spike GRB must be to satisfy a trigger bin size equal 
to the GRB duration, the ``best case" for detectability.  

BATSE's finite size also limits the duration of GRBs
it can detect above any given peak flux.   Assuming
a negligible background, the limiting peak flux is
 \begin{equation}
   P_{lim} \sim {N \over A \ t_{dur}} ,
 \end{equation}
where $P_{lim}$ is the limiting peak flux a detector
of cross-sectional area $A$ can measure, and $N$ is 
the number of photons needed for a significant detection.
For BATSE in Figure 1, we assume $N \sim 10$ and 
$A \sim 10^4$ cm$^2$.  This detector-size limit is shown by
the dotted line in Figure 1.  

Figure 2 shows a histogram of durations for BATSE 
3B GRBs.  Only those GRBs in the BATSE 3B 
with listed $T_{50}$s, and peak flux on the 64 ms 
scale above 2 photons cm$^{-2}$ sec$^{-1}$ are 
plotted in the unfilled histogram region: 295 GRBs 
in all.  BATSE is expected to be about 
98 percent complete to this peak flux level 
(Fishman et al. 1994a).  All rates plotted in Figure 2 are 
normalized to the BATSE measured rate for GRBs 
with $T_{50}$ durations between 8.192 and 16.384 
seconds and peak fluxes above 2 photons cm$^{-
2}$ sec$^{-1}$.  Bursts with these durations have
the highest rate recorded by BATSE.

Superimposed on BATSE's 3B detections at short 
durations is a cross-hatched histogram.  This region 
represents an upper limit on the rate that GRBs 
might have been measured, were BATSE's 
sensitivity level somehow held constant below 64 ms.  
More precisely, GRBs falling in this region fall below a 
1-$\sigma$ upper limit estimate of how many short 
GRBs would have been measured to a peak flux 
limit on a 64 ms time-scale where BATSE is about 98 \% 
complete. The 1-$\sigma$ lower limit was near zero 
for most of these short time bins, so it is also 
possible that very few GRBs occur in this region.  

This extrapolation to short GRBs 
was computed as follows.  Candidate rate levels 
were assumed for GRBs with durations between 
1 ms and 64 ms, from which a virtual catalog of 
GRBs was created.  The detection rate in each 
duration bin was confined 
to the measured, differential, BATSE 3B log N - 
log P distribution.  
This virtual GRB catalog was then subjected to 
trigger thresholds similar to BATSE (Nemiroff et al. 1996).  
Specifically, 
GRBs with peak flux falling below equation (1) 
were cut, leaving a duration histogram that was 
compared to a 1-$\sigma$ upper limit on the base 
BATSE 3B histogram (unhatched region) of Figure 
2.  The candidate rate levels were iterated until a 
satisfactory match was found.  The virtual GRB 
catalog was made much larger than the actual GRB 
catalog to minimize numerical error.  Between  
1 $\mu$s and 1 ms, extrapolation of the bright 
portion of the GRB log N - log P indicates that as
duration drops by a factor of 2, the acceptable
1-$\sigma$ rate should increase by a factor of  
roughly $2^{1.5}$.

\section{The Abundance and Utility of GRB Spikes}

Inspection of the hatched short duration part of the 
Figure 2 histogram indicates that a significant rate 
of short GRBs might exist but go undetected by 
BATSE.  In fact, to a given peak flux level, 
there may be more GRBs that occur 
between 1 and 2 ms than between 8.192 and 16.384 
seconds.  The limit on the rate of microsecond spikes
is even less well constrained, and could even be
an order of magnitude greater. 
Alternatively, these ``spike" GRBs might not 
exist at all: both are acceptable to within 
1-$\sigma$ errors.  Were spike GRBs to exist, 
however, they would not be recoverable by analysis 
on existing BATSE archival data because the 
highest time resolution, continuously recorded 
BATSE data type has 1.024 sec bins (``DISCLA" 
data).  

Currently there are claims that a non-negligible 
fraction of our Galaxy is composed of compact 
objects on the order of a fraction of a solar mass 
(see, for example, Alcock et al 1996).  Also, a 
non-negligible fraction of the baryon content of our 
universe, on order $\Omega \sim 0.01$, might exist 
in a stellar or stellar remnant form (see, for 
example, Carr 1994, and references therein).  If so, 
a sufficiently well-tuned GRB detector might resolve 
these microlenses temporally. The 
precise relation for a single standard compact 
gravitational microlens is (Krauss and Small 1991;
Mao 1992; Nemiroff et al. 1993)
\begin{equation}
 \Delta t = { R_S (1+z_L) \over c }
 \left( {{f - 1 \over \sqrt{f} } - {\rm ln} f
          } \right) ,
 \end{equation}
where $\Delta t$ is the time delay between the two 
bright images, $R_S$ is the 
Schwarzschild radius of the lens ($R_S \sim 3$ km 
$M/M_{\odot}$), $z_L$ is the redshift of the lens, 
$f$ is the dynamic range between the two bright
images ($f>1$), and $c$ is the speed of light.   For $f = 6.85$, 
the dynamic range when a lens is displaced one Einstein
ring from the source,  $z_L =$ 0.5 and lens mass of 0.5 
$M_{\odot}$, the resulting time delay between 
images is $\Delta t \sim$ 2.3 microseconds.
Significantly more disparate images would bring significantly
longer time delays.

How often would microlensing occur?  For sources 
at cosmological redshift near unity, the expected 
microlensing rate is on order $\Omega_{lens}/8$ 
for an $\Omega = 1$ universe
(see, for example, Nemiroff 1989).  For GRBs at a 
canonical cosmological redshift of unity and the 
known star field estimate of $\Omega_{stars}$ of 
0.002 (Carr 1994), roughly one in 4000 GRBs 
would be expected to undergo detectable 
microlensing. For MACHOs of 0.5 $M_{\odot}$ 
composing half of galaxy halos (see, for example, 
Alcock et al. 1996), the universal density would be 
quite uncertain owing to our lack of knowledge of 
the true extent of spiral galaxies.  However, were
$\Omega_{MACHO} \sim$ 0.1, this would result 
in one in 110 GRBs undergoing detectable 
microlensing.  From this one can see that even a 
null result in a microlensing search in short GRBs  
could indicate the extent to which MACHOs 
populate galactic halos.

Even higher microlensing rates would be expected 
for even less popular cosmological scenarios, 
possibly involving a cosmological constant and/or a 
significantly higher $\Omega_{lens}$.  Of course, 
the probability of any specific GRB undergoing 
microlensing is a strong function of the GRB 
redshift and the detectable dynamic range, 
although {\it not} a function of lens mass 
(Nemiroff 1989).  

GRBs of all durations undergo gravitational 
microlensing, but detector size, background, and
long duration bin sizes have made 
this effect, so far, undetectable.  Bursts at or below a  
$\mu$s in duration would likely show microlensing 
with an easily discerned double peaked light curve 
signature.  Longer duration bursts with significant 
temporal variability at or below a $\mu$s might 
also show detectable microlensing, although it 
might not appear so obvious to the untrained observer.
For this reason, we refer sometimes
to ``GRB spikes" instead of ``spike GRBs" so as to
more clearly incorporate the possibility that 
longer duration GRBs might have internal spikes that 
could show microlensing.  In an informal search
through BATSE TTE data, the shortest significant 
fluctuation we were able to find was on the order of 
100 $\mu$s.  Indeed, Schaefer et al. (1993) have 
shown that bright BATSE GRBs are not, in general, 
a superposition of bright microsecond flares 
(Mitrofanov  1989).

As an aside, we note that these results 
might have some bearing on past and 
future time dilation estimates for GRBs.  First, one 
must be very careful about including short bursts of 
low peak flux.  For GRBs with $T_{50}$s below 
about 2 seconds, BATSE trigger limits create an 
artificial dearth of low peak flux GRBs.  Were this 
unaccounted for, one's analysis might be skewed by 
a false dim/long association (for more discussion 
see Lee \& Petrosian 1996).   We note that the 
Norris collaboration searches for time dilation 
(see, for example, Norris et al. 1994 \& Norris et al. 
1995) routinely exclude bursts this short in their 
time dilation analysis.

\section{Discussion and Conclusions }

Several assumptions went into the above 
extrapolated GRB rate estimates.  One 
was that the log N - log P distribution for GRBs
is not a strong function of duration.  A 
popular special case of this occurs when 
instantaneous GRB peak flux is a true standard candle.
Were fluence a better standard candle for GRBs,
however, (see, for example, Petrosian \& Lee 1996)
shorter GRBs at a given true peak flux 
would be increasingly nearby, and hence at 
decreasing  probability of undergoing detectable 
microlensing.

Another implicit assumption is that the correct
GRB physical model does not prohibit GRB spikes. 
Given the abundance of suggested physical models
(see, for example, Nemiroff 1994), we know of no 
fundamental reason why   
GRB spikes should not exist.  Even if GRB 
timescales were somehow related to the light crossing time
of a neutron star, on order 100 $\mu$s, 
emission might result from the collision of relativistic 
shells with $\gamma$ factors in the hundreds, 
perhaps providing the needed time structure.
If spike or short GRBs never occurred, this 
would be interesting 
by itself, as it could be used to indicate 
a size scale for the emitting region, 
and hence give a clue to the 
emission mechanism behind GRBs.

Spike GRBs are even predicted for some models,
including small evaporating black holes (Cline et al. 1997).
In this case, lack of detection limits the cosmological
density of these objects (Fichtel et al. 1994).

Inspection of Figure 2 indicates that some short GRBs with 
durations below 64 ms may be theoretically detectable 
by BATSE at a rate exceeding that in the 3B catalog, 
although not 
recoverable from existing archival data.  Toward 
this end, BATSE's single sweep pulsar data type 
can be programmed to run for approximately
($t_{bin}/$ 32 ms) fraction of the time (Meegan 1996).  
Such a program would also better
determine the rate of relatively 
dim Terrestrial Gamma Flashes, events with
typical durations near 10 ms (Fishman et al. 1994b; 
Nemiroff, Bonnell \& Norris 1997).

Future GRB detectors may be designed to 
explore this important spike
regime.  Specific attributes desirable in future 
detectors are larger area, smaller time bin 
sizes, sensitivity to photons of lower energy, and faster triggers.  
Inverting and augmenting Eq. 3 we see that
\begin{equation}
  A \sim \left( {N \over 1 \ {\rm photon} } \right)
 \left( {1 \ {\rm ph \ \ cm^{-2} \ sec^{-1}} \over P_{lim} } \right)
 \left( {1 \ {\rm sec} \over t_{dur} } \right) 
 \left(  E \over { 50 \ {\rm keV} } \right)^2 , 
\end{equation}
where $E$ is the lowest energy the detector can efficiently
measure.  The last factor derives from a canonical estimate for the
spectrum of GRBs which yields significantly more 
photons at lower energies.  To detect 10 photons from a GRB of duration 
1 $\mu$s at the BATSE minimum peak flux of 
1 photon cm$^{-2}$ sec$^{-1}$ above 50 keV, 
the detector area must be on order 10$^7$ cm$^2$, 
or about 36 meters in diameter.  Although this is prohibitively large, 
detectors sensitive to lower energy photons are
more reasonably sized, with a 5 keV detector 
sensitive to GRB spikes at 3.6 meters in diameter, 
and a 1 keV detector sensitive to GRB spikes can
be less than one meter in diameter. 

At low energies, below about 10 keV, curvature in GRB
spectra might make equation 5 optimistic and hence overestimate 
the ease of detections.  Also, the increasing background 
toward lower energies makes detection
more difficult for wide field instruments.  
Such detectors must be properly shielded to reduce 
the background.   An additional confounding factor might
arise were GRB spikes intrinsically longer at lower energies.  
At higher energies, 
we note that EGRET was used to 
search for GRB spikes at 250 MeV at a level of about
10$^{-3}$ photons cm$^{-2}$ sec$^{-1}$, 
without success (Fichtel et al. 1994).

In sum, were cosmological gamma-ray bursts (GRBs) 
with duration or variability of the order of a microsecond
to exist (``GRB spikes"), they would 
time-resolve intervening stars as cosmic microlenses.  
Given the BATSE 3B catalog and standard 
canonical assumptions, we find to one-sigma
accuracy that the rate of GRBs with durations 
at and below 1 ms 
could be higher than for GRBs of any other 
duration, or could be zero. For Omega of unity,
MACHOs comprising one tenth of the critical density
would be expected to be detected as microlenses in roughly
one in 110 GRB spikes.  
From the known star field, however, only one in 4000
GRB spikes would undergo detectable microlensing.
Although present spacecraft detectors are too small 
to measure a significant GRB spike rate, 
future detectors optimized for size, energy, and memory
might constrain or detect an abundance
of baryons, stars, and MACHOs in the universe.

We acknowledge helpful comments 
by C. Ftaclas, C. Kouveliotou,  C. Meegan, B.  Paczynski, J. Scargle, and 
the anonymous referee.  This research was
supported, in part, by grants from NASA and NSF.

\clearpage

\figcaption[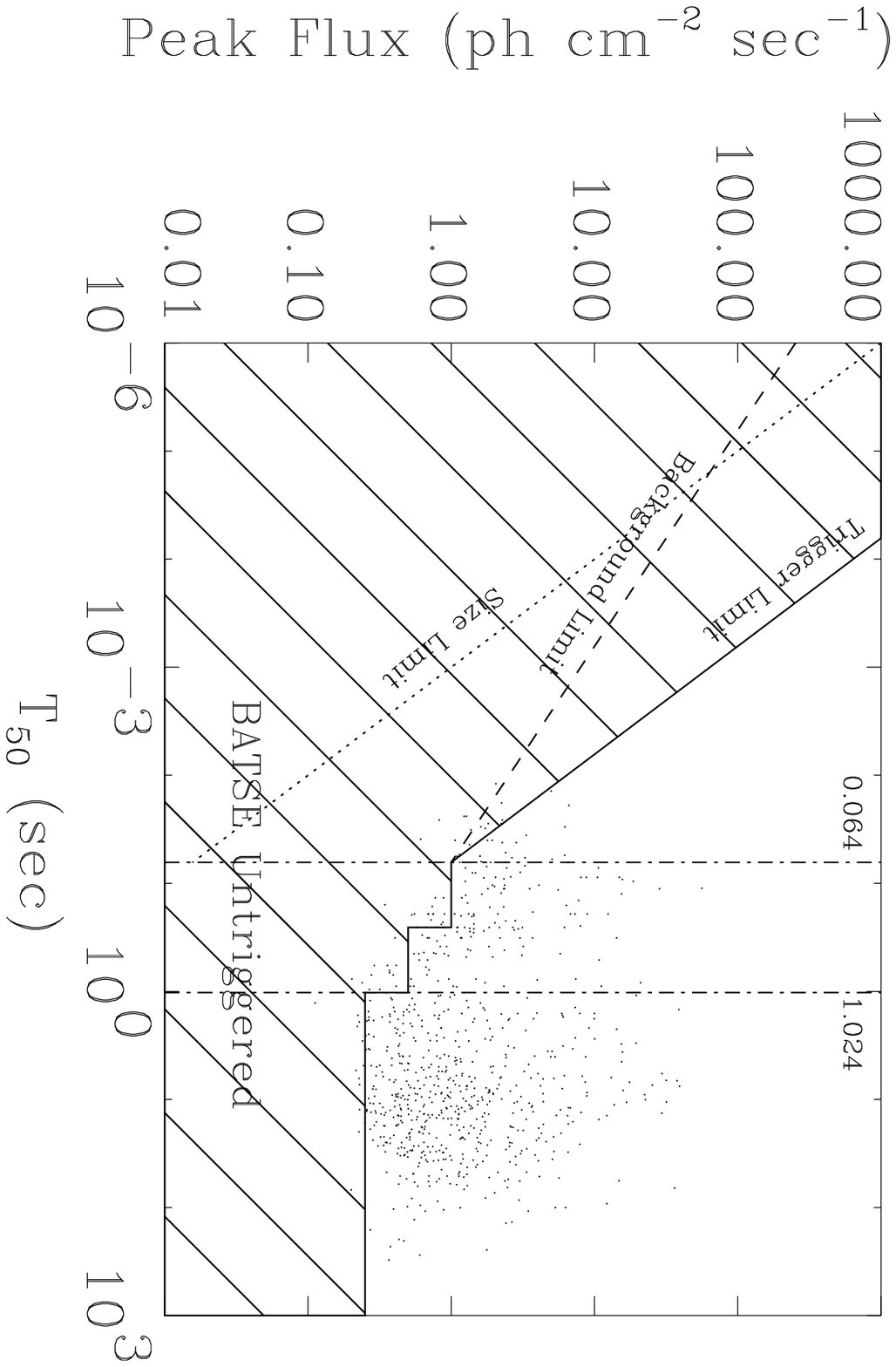]{A plot of peak flux versus 
duration for BATSE 3B GRBs. GRBs 
occurring in the hatched region with lines running 
from the lower left to the upper right would not, in 
general, trigger BATSE.   GRBs below the dashed line
would be hidden below BATSE's background, and
GRBs below the dotted line would deposit less
than 10 photons on BATSE.
\label{fig1}}

\figcaption[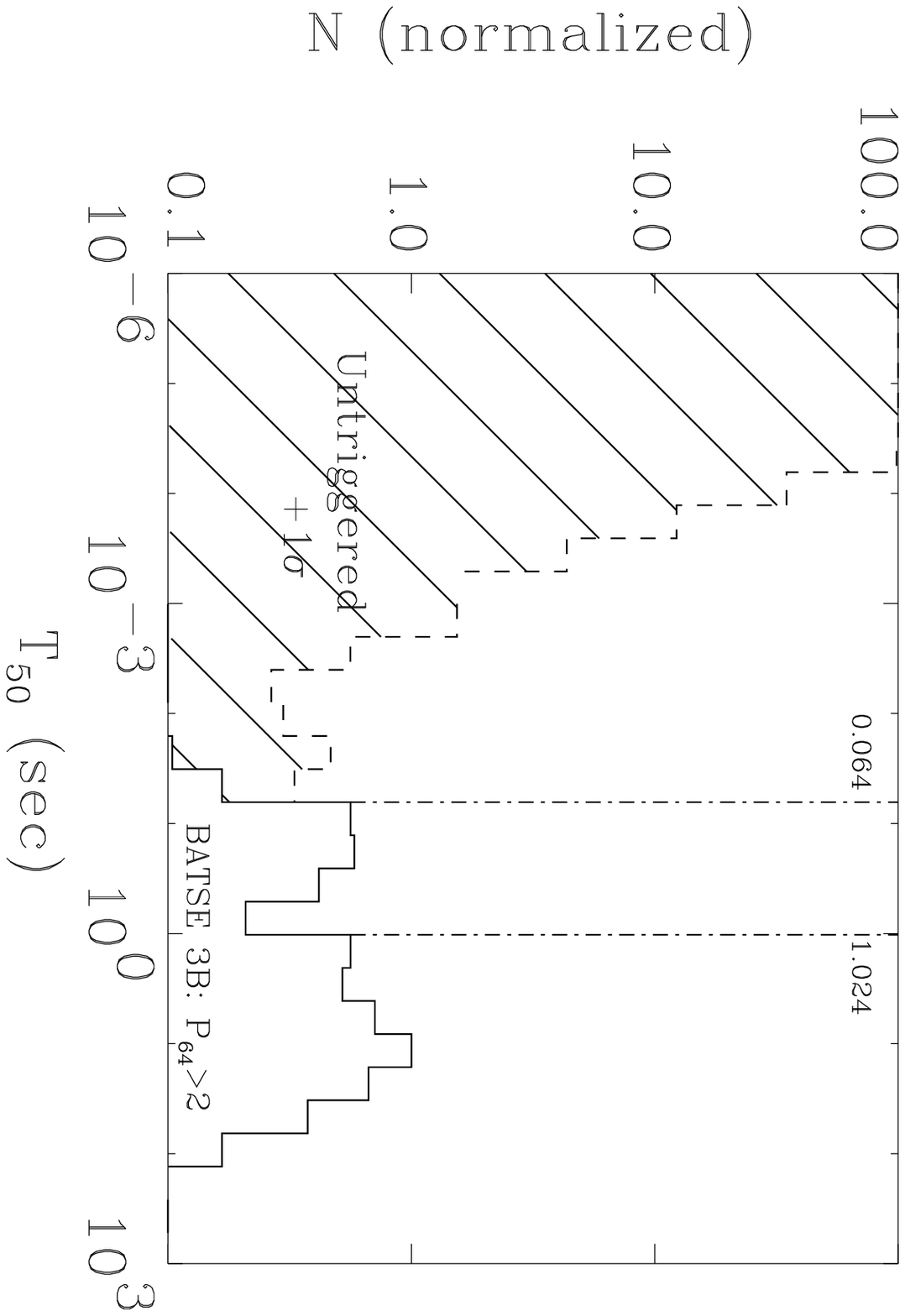]{A duration histogram.  The 
unhatched region results from GRBs in the BATSE 
3B catalog.  Burst rate is normalized to the BATSE 
bin covering durations between 8.192 and 16.384 
seconds.  Bursts occurring in the hatched region 
with lines running from the lower left to the upper 
right are one-sigma upper limits on the rate for 
GRBs with peak flux greater than 2 photons per 
square centimeter per second.  Note that within one sigma
upper limits, the rate of GRBs with durations below
1 microsecond may exceed the GRB rate at 10 seconds.
\label{fig2}}

\clearpage
\plotone{spikef1.eps}

\clearpage
\plotone{spikef2.eps}

\end{document}